\documentclass[letterpaper,pdftex,rmp,twocolumn,amsmath,amssymb,groupedaddress,floatfix]{revtex4}
\usepackage{graphicx}
\RequirePackage[sort&compress]{natbib}
\usepackage{natbib}
\usepackage{xcolor}
\usepackage{textcomp}
\usepackage{ulem}
\usepackage{makecell}
\usepackage{float}
\usepackage{setspace}
\usepackage[a]{esvect}
\usepackage[font=sf,size=footnotesize,justification=RaggedRight,singlelinecheck=false]{caption}
\usepackage[sf,bf,small]{titlesec}
\titlespacing*{\section}{0pt}{1em}{0em}

\usepackage{mathrsfs}
\usepackage{amsmath,amssymb}
\usepackage{bm}

\definecolor{darkgray}{rgb}{0.25,0.25,0.25}
\definecolor{darkred}{rgb}{0.89,0.10,0.11}
\definecolor{darkblue}{rgb}{0.12,0.39,0.62}
\usepackage{url}

\usepackage[pdftex,breaklinks=true,colorlinks=true,citecolor=black,linkcolor=black,menucolor=black,urlcolor=darkblue,pdfborder={1 0 0}]{hyperref}
\hypersetup{pdftitle={Respondent-driven sampling bias induced by clustering and community structure in social networks},pdfauthor={L.E.C. Rocha (2015)}}
\bibpunct{}{}{,}{s}{,}{,}

\usepackage{booktabs}
\usepackage{color}

\begin{document}
\makeatletter
\renewcommand\@biblabel[1]{#1.}
\makeatother

\newcommand{\dif}{\mathrm{d}}

\newcommand{\lr}[1]{\textcolor{red}{#1}}

\renewcommand{\figurename}{Figure}
\renewcommand{\thefigure}{\arabic{figure}}
\renewcommand{\tablename}{Table}
\renewcommand{\thetable}{\arabic{table}}
\renewcommand{\refname}{\large References}

\addtolength{\textheight}{1cm}
\addtolength{\textwidth}{1cm}
\addtolength{\hoffset}{-0.5cm}

\setlength{\belowcaptionskip}{1ex}
\setlength{\textfloatsep}{2ex}
\setlength{\dbltextfloatsep}{2ex}

\title{Respondent-driven sampling bias induced by clustering and community structure in social networks}

\date{\today}

\author{Luis E C Rocha}
\email{luis.rocha@ki.se}
\affiliation{
Department of Public Health Sciences, Karolinska Institutet, Stockholm, Sweden \\
Department of Mathematics and naXys, Universit\'e de Namur, Namur, Belgium}

\author{Anna E. Thorson}
\affiliation{Department of Public Health Sciences, Karolinska Institutet, Stockholm, Sweden}

\author{Renaud Lambiotte}
\affiliation{Department of Mathematics and naXys, Universit\'e de Namur, Namur, Belgium}

\author{Fredrik Liljeros}
\affiliation{Department of Sociology, Stockholm University, Stockholm, Sweden}

\begin{abstract}
Sampling hidden populations is particularly challenging using standard sampling methods mainly because of the lack of a sampling frame. Respondent-driven sampling (RDS) is an alternative methodology that exploits the social contacts between peers to reach and weight individuals in these hard-to-reach populations. It is a snowball sampling procedure where the weight of the respondents is adjusted for the likelihood of being sampled due to differences in the number of contacts. In RDS, the structure of the social contacts thus defines the sampling process and affects its coverage, for instance by constraining the sampling within a sub-region of the network. In this paper we study the bias induced by network structures such as social triangles, community structure, and heterogeneities in the number of contacts, in the recruitment trees and in the RDS estimator. We simulate different scenarios of network structures and response-rates to study the potential biases one may expect in real settings. We find that the prevalence of the estimated variable is associated with the size of the network community to which the individual belongs. Furthermore, we observe that low-degree nodes may be under-sampled in certain situations if the sample and the network are of similar size. Finally, we also show that low response-rates lead to reasonably accurate average estimates of the prevalence but generate relatively large biases.
\end{abstract}

\maketitle

\noindent

\section{\large Introduction}

In order to estimate the prevalence of diseases, traits or behaviors in particular social groups or even in the entire society, researchers typically rely on samples of the target population. A carefully selected sample may generate satisfactory low standard errors with a bonus of optimizing research resources and time. A common challenge is to obtain a significant and unbiased sample of the target population. This is particularly difficult if this population of interest is somehow segregated, stigmatized, or in some other way difficult to reach such that a sampling frame cannot be well defined. These so-called hidden (or hard-to-reach) populations may be for example man-who-have-sex-with-man (MSM), sex-workers, injecting drug users, criminals, homeless, or minority groups~\cite{Sudman1988}.

In 1997, Heckathorn introduced a new methodology to sample hidden populations named respondent-driven sampling (RDS)~\cite{Heckathorn1997}. RDS exploits the underlying social network structure in order to reach the target population through the participants' own peers. The method consists in a variation of the snowball sampling where the statistical estimators have weights to compensate the non-random nature of the recruiting process, i.e.\ that individuals with many potential recruiters have a higher chance to be sampled. In RDS, researchers select seeds to start the recruitment. A seed person then invites a number of other individuals to participate in the survey by passing a coupon to them. Those successfully recruited respond a survey and get new coupons to invite a number of other individuals within their own social network, and the process is repeated until enough participants are recruited. Successful recruitment and participation in the survey are both financially compensated. A fundamental assumption is that each participant knows the number of his or her own acquaintances in the target population, or in the network jargon, his or her own degree. This information is used as weights to estimate the prevalence of the variable of interest in the study population.

\begin{figure*}[ht]
\centering
\includegraphics[scale=1.0]{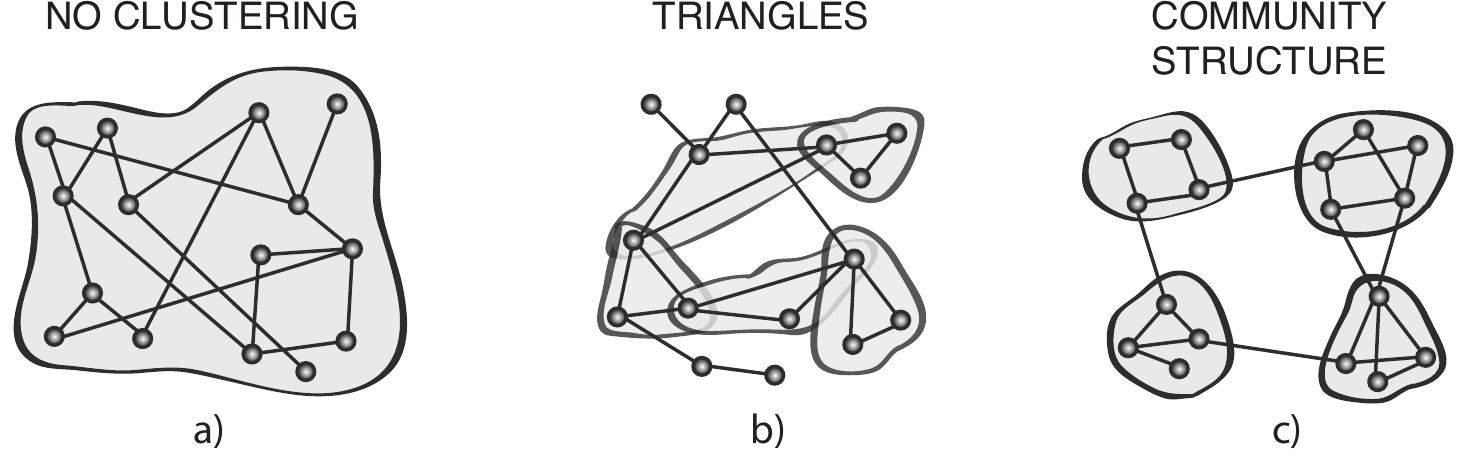}
\caption{\textbf{Networks with and without structure.} The panel shows a schematic network a) completely random, i.e.\ without triangles and community structure, b) with 5 triangles and without community structure, and c) with 4 network communities and 1 triangle (in the bottom-right community). }
\label{fig:01}
\end{figure*}

The perhaps most popular RDS statistical estimator is due to Volz and Heckathorn, who devised a Markov process whose equilibrium distribution is the same as the distribution of the target population~\cite{Volz2008}. This estimator is derived after a series of assumptions regarding both the underlying network structure and the recruitment process \textit{per se}. Although the assumptions are generally reasonable, sometimes they are relatively strict for realistic settings, as for example, the uniformly random selection of peers, persistent successful recruitment, and sampling with replacement~\cite{Semaan2010}. These and other assumptions have been scrutinized in previous theoretical studies and the estimator has performed satisfactory in different scenarios using both synthetic~\cite{Abdul2006, Salganik2006, Gile2010} and real networks~\cite{Lu2012, Verdery2014}. A number of real life studies have also concluded that RDS is an effective sampling method for various categories of hidden populations (See for example Refs.~\cite{McKnight2006, Robinson2006, Abdul2006, Abramovitz2009, Iguchi2009}).

Social networks are however highly heterogeneous in the sense that the structure of connections cannot be represented by characteristic values. This is the case of the number of contacts per individual or of the level of clustering between them~\cite{Newman2010, Costa2011}. Since the RDS dynamics is constrained by the network structure, one may expect that different patterns of connectivity affect the recruitment chains. For example, the network structure may be such that a recruitment tree grows only in one part of the network~\cite{Martin2003, Burt2010, McCreesh2011}. In realistic settings using sampling without replacement, even if all individuals are willing to participate, trees may simply die out because a network has been locally exhausted and bridging nodes block further propagation of coupons to other parts of the network~\cite{Johnston2013}. Such situation is not unlikely in highly clustered sub-populations where coupons may simply move around the same group of people. Previous theoretical studies have addressed some of these network constrains by studying the RDS performance on either synthetic structures~\cite{Salganik2006, Gile2010} or samples of real networks~\cite{Lu2012, Verdery2014}. Each approach to model social networks has its own advantages and limitations. On one hand, simple synthetic structures and sampling processes are unrealistic but allows some mathematical treatability and thus intuitive understanding. On the other hand, samples of real networks may suffer biases themselves due to their own sampling and thus potential incompleteness of data~\cite{Lee2006, Latapy2008}.

Network clustering is particularly important in the context of social networks and should be carefully assessed. It may have different meanings but here we associate clustering to social triangles, i.e.\ the fact that common contacts of a person are also in contact themselves. Network communities are also a form of clustering in which groups of individuals are more connected between themselves than with individuals in other groups. As already mentioned, clustering in all its forms is not uniform across a network. In practice, it means that one may find hidden sub-populations within the study population. Examples include social groups with particular features (e.g.\ wealth, foreigners, ethnic minorities) embedded in the target population~\cite{Johnston2013}, transsexuals in populations of MSM, or geographically sparse populations~\cite{Burt2010}. While these sub-populations may potentially be removed by defining a more strict sampling frame, social groups (or communities) are inherent of social and other human contact networks~\cite{Wasserman94, Costa2011}. Note that network clustering is not the same as homophily, that is the tendency of similar individuals to associate, but one may enhance the other. For example, individuals may share social contacts because they live geographically close, share workplaces, or are structured in organizations (potentially leading to network clustering) but may be completely different in other aspects (low homophily in wealth, health status, gender, infection status, and so on).

In this paper, we use computational algorithms to generate synthetic networks with various levels of clustering and with network communities of various sizes, aiming to reproduce structures observed in real social networks. Using realistic parameters, we simulate a RDS process using these networks and quantify the performance of the RDS estimator in different scenarios of the prevalence of an arbitrary variable of interest. The paper is organized such that we first analyze how triangles and community structure affect how the RDS spread in the network when it comes to size of transmission trees and generation of recruitment. Then we investigate how clustering affects the validity and reliability of the RDSII estimator as a function of different willingness to participate (response-rates) in the population. We also test the effect of clustering for scenarios where the variable under study is correlated with the degree of the nodes and the size of the network community. Thereafter, we study the consequences of the biased selection of seeds, the bias induced by network structure in samples of real social networks, and the effect of restarting the seeds during the sampling experiment.

\section{\large Materials and Methods}
\noindent

We describe in this section the models used to generate the synthetic networks with different number of triangles and varying levels of community structure, the empirical networks, the model to simulate the RDS dynamics, the protocols to artificially distribute the infections in the target population, and the estimator and other statistics used for the analysis.

\subsection{Study networks} 
\label{Methods_A}

A social network is defined by a set of nodes representing the population and a set of links representing the social contacts, as for example acquaintances or friendship, between two individuals. The network structure can be characterized by different network quantities~\cite{Wasserman94,Newman2010}. The most fundamental quantity is the degree $k$ that represents the number of links of a node or equivalently the number of contacts of an individual. The assortativity of a network measures the tendency of nodes with similar degree to be connected. The number of triangles and the clustering coefficient are used to measure the local clustering in the network. A triangle corresponds to the situation where two contacts of a node are also in contact themselves, and the clustering coefficient is a normalized count of the number of triangles. A network community, on the other hand, is a group of nodes that are more connected between themselves than with nodes of other groups. A fundamental property of the network community structure is that only a few nodes link (or bridge) different communities, these nodes are also known as bottlenecks because they constrain the diffusion, or the sampling process, in the network. If there are only a few bridging nodes, one says that the community structure is strong, whereas many bridging nodes weaken the community structure reducing the bottlenecks between groups.

\subsubsection{Synthetic Networks} We use computational algorithms able to generate synthetic networks with tunable number of triangles (Fig.~\ref{fig:01}b) or of community structure (Fig.~\ref{fig:01}c). These algorithms are not expected to reproduce a particular social network but to generate various structures observed in social networks more realistically than previously studied structures~\cite{Salganik2006}. Our reference random network is obtained by simply connecting pairs of nodes for a given degree sequence, a procedure that results on a negligible number of triangles and no network community structure (Fig.~\ref{fig:01}a). This model is also know as the configuration model~\cite{Newman2010}.

The first algorithm, due to Serrano and Bogu\~na~\cite{Serrano2005}, generates networks with a varying number of triangles and assortativity. In this algorithm, an \textit{a priori} degree sequence is chosen following a given distribution $P(k)$ of node degree $k$. We choose a power-law degree distribution with a small exponential cutoff, i.e.\ $P(k) \propto k^{-2.5}\exp(-0.0001k)$. If no or very small costs are associated with keeping links alive, scale-free distributions are reasonable models for empirical distribution, otherwise we usually observe broad scale distributions not necessarily power-law-like. Generally speaking, this degree distribution is thus not expected to be the most appropriate distribution of contacts in real populations but it captures the right-skewed degree heterogeneity typically observed in social groups~\cite{Newman2010, Costa2011, Lu2012}. This heterogeneity means that the majority of nodes has only a few contacts whereas a small number of them has several contacts. We fix the minimum possible degree to $x_{min}=3$ in order to obtain an average degree $\langle k \rangle \sim 7$. Furthermore, an \textit{a priori} clustering coefficient is chosen such that a given number of triangles is defined for each degree class $k$. The algorithm evolves by randomly selecting three different nodes and forming a triangle between them, respecting the distribution of triangles per degree class. As soon as no new triangles can be formed, the remaining links are uniformly connected (i.e.\ the configuration model) such that no links are left unconnected. Self-links are forbidden. A parameter $\beta$ controls the assortativity (assortativity increases with decreasing $\beta$) and the parameters $c_0$ and $\alpha$ control the expected clustering coefficient (clustering increases with increasing $c_0$ and decreases with increasing $\alpha$). In this paper, we use $c_0=0.5$, $\alpha=0.3$ and $\beta=1.0$ (for the configuration with many triangles) and $c_0=0.5$, $\alpha=1.0$ and $\beta=1.0$ (for the configuration with few triangles).

The second algorithm, developed by Lancichinetti and Fortunato~\cite{Lancichinetti2009}, is used to create networks with community structure. Here one starts by choosing the distribution of degrees and the distribution of community sizes. In both cases, we use power-law distributions to capture the heterogeneity in the degree and in the community size as observed in some real social networks~\cite{Newman2010, Costa2011}. Other choices of probability distribution may be more suitable for specific populations but here again we want to study the heterogeneity in the community sizes. The degree distribution has the same parameters as used in the first algorithm, the power-law distribution of community sizes has exponent $-1$ and community sizes are limited between $10$ and $1000$ nodes. These values are chosen to guarantee that a sufficient number of communities are large in size and at the same time, enough small-sized communities are represented. For example, higher values of the exponent would result in relatively more small-sized communities. These values are also constrained by the number of links and the level of overlapping of communities (see below), and are chosen to generate a network with a single connected component. The number of overlapping nodes and the number of communities that each node belongs to are inputs of the algorithm. Overlapping means that a number of nodes belong to more than one community (these are the bridging nodes) while the rest of the nodes only belong to single communities. One may further select a mixing parameter $\mu$ to add random links between the bridge nodes and randomly chosen communities (to weaken the community structure). Therefore, small overlapping and small mixing generate stronger community structures. We set $\mu=0$, and select $100$ or $1000$ overlapping nodes in $5$ communities respectively for strong and strong-moderate community structures. For moderate-weak and weak community structures, we set respectively $\mu=0.3$, and $100$ and $1000$ overlapping nodes (in $5$ communities as well).

For each algorithm, to obtain the statistics, we generate $10$ versions of the network with the same set of parameters and with $10000$ nodes each, which is also the size of the target or study population. 

\subsubsection{Empirical Networks} We also study RDS using real-life networks. We perform simulations on 5 samples of empirical contact networks representing different forms of human social relations. Three data sets correspond to email communication, two between members of two distinct universities in Europe (EMA1~\cite{Guimera2003}, EMA2~\cite{Eckmann04}) and one between employees of a company (ENR)~\cite{Enron2009}. In these datasets, nodes correspond to people and social ties are formed between those who have sent or received at least one email during a given time interval. One data set corresponds to friendship ties between US high-school students (ADH)~\cite{Moody2001}. The last data set corresponds to online communication between members of an online dating site (POK)~\cite{Holme04}. Similarly to the email networks, if two members have exchanged a message through the online community, a link is made between the respective nodes. Although some of these data sets do not correspond to social networks in which RDS would take place, they serve as realistic settings capturing the network structure of actual social relations. We have selected data sets with diverse sample sizes and network structure in order to cover various contexts and configurations (Table~\ref{tab:01}).

\begin{table}[htb]
\centering
\begin{tabular}{cccccc}
\hline
        & EMA1 & ADH & EMA2 & POK & ENR \\
\hline
$N$ & 1,133 & 2,539 & 3,186 & 28,295 & 36,692 \\
$E$ & 5,451 & 10,455 & 31,856 & 115,335 & 183,831 \\
$cc$ & 0.22 & 0.15 & 0.26 & 0.05 & 0.50 \\
$C$ & 57 & 200 & 71 & 2,615 & 2,441 \\
$C_S$ & 2 & 1 & 1 & 1 & 1 \\
$C_L$ & 151 & 222 & 1,205 & 2,621 & 1,481 \\
\hline
\end{tabular}
\caption{\textbf{Summary statistics of the empirical networks used in this study.} Number of nodes ($N$); number of links ($E$); clustering coefficient $cc$; number of communities $C$; size of the smallest community $C_S$ and size of the largest community $C_L$, according to the MapEquation algorithm~\cite{Rosvall2008}. }
\label{tab:01}
\end{table}

\subsection{RDS model}
\label{Methods_B}

We simulate the sampling by using a stochastic process reproducing several features of a realistic RDS dynamics. Our model further adds a continuous-time framework and the response-rate can be controlled. We use similar parameters as typically used in the literature~\cite{Lu2012, Gile2015}.

We start by uniformly selecting (unless otherwise stated) $10$ random nodes as seeds for the recruitment. After a time $t$, sampled from an exponential distribution, each seed chooses uniformly three of its contacts and pass one coupon to each of them. The exponential distribution is chosen because in our model we assume that the recruitment follows a Poisson process. We select the average waiting time to be $5$, meaning that a node waits on average $5$ time steps (e.g.\ 5 days) before inviting its contacts. Therefore, after waiting $t$ time steps, and with probability $p$, that represents the probability of participation (or response-rate, i.e.\ one minus the probability of not returning a coupon), each of these contacts recruits three of their own contacts that have not participated yet (sampling without replacement). If a node accepts to invite its own contacts (i.e.\ accepts to participate), we add this node in the sample. The process continues until all possibilities of new recruitments are exhausted or, at maximum, when a specific sample size is reached. Note that this continuous-time model is equivalent to a discrete-time model in which randomly chosen nodes update their status sequentially. We assume that if a node refuses to participate once, it becomes available for recruitment by other nodes as if it was never invited. We repeat the simulation of the RDS dynamics $50$ times for each synthetic network and $500$ times for each empirical network.

\begin{figure*}[htb]
\centering
\includegraphics[scale=1.0]{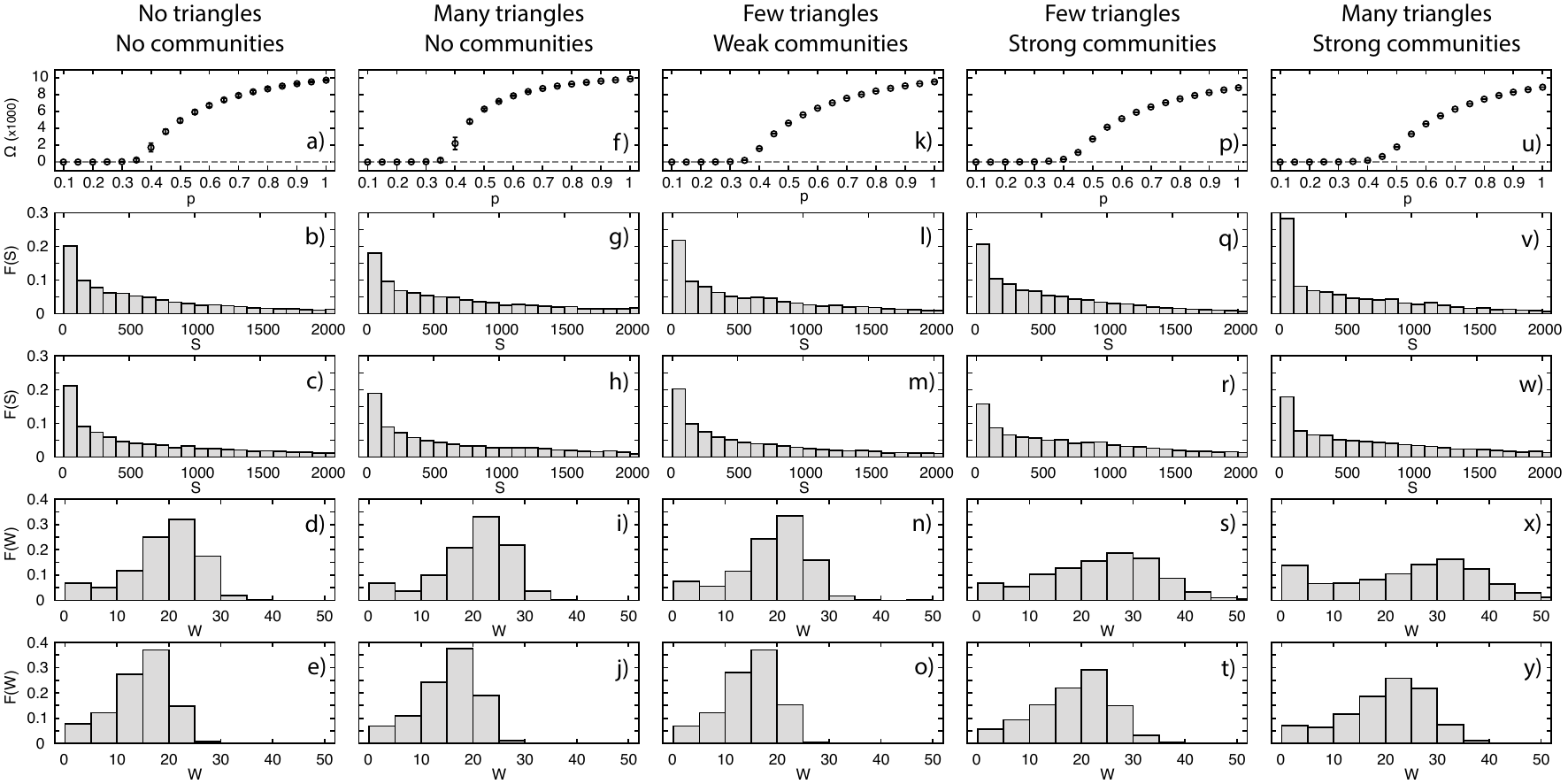}
\caption{\textbf{Statistics of recruitment trees.} The panel shows a,f,k,p,u) the total number of recruited subjects $\Omega$ for different response-rates $p$ (dotted lines correspond to zero, bars on points correspond to the standard error), the distribution of size of recruitment trees $S$ per seed for response-rates b,g,l,q,v) $p=0.7$ and c,h,m,r,w) $p=1.0$ (histogram bin size is 100), and the distribution of number of waves $W$ per seed for response-rates d,i,n,s,x) $p=0.7$ and e,j,o,t,y) $p=1.0$ (histogram bin size is 5). The underlying structures are random networks with different number of triangles and different levels of community structure (See Section~\ref{Methods_A}). }
\label{fig:02}
\end{figure*}

\subsection{Prevalence of the study variable} 
\label{Methods_C}

In RDS studies, one is interested in quantifying the prevalence of some variable $A$ in the target population. This variable may represent, for instance, being tested positive for a given disease, being male or female, the ethnicity, or having a particular physical trait. In this paper, to simplify the notation, we say that an individual and its respective node is infected with $A$ or not infected with $A$. We use different protocols to infect a fraction of $25\%$ of the network nodes with the quantity $A$. The remaining nodes are thus assumed to be non-infected.

The reference case (RI) corresponds to uniformly selecting the nodes within the target population, i.e.\ the infection $A$ is uniformly distributed in the network.

The preferential case (PI) corresponds to selecting nodes in decreasing order of degree, i.e.\ we start at nodes with the highest degree and infect them with $A$ until $25\%$ of the nodes become infected. To add some noise (case PRI), we select $20\%$ of the infected nodes, cure them, and redistribute these infections uniformly in the network such that the total number of infected nodes remains fixed.

The other two cases consist on infecting nodes according to the community structure. In the first case (SI), we initially infect nodes in the smallest communities until $25\%$ of the nodes become infected. In the second case (BI), we infect nodes in the largest communities until the same fraction of $25\%$ of nodes get infected. To reduce homophily, we add noise by selecting $40\%$ of the infected nodes, curing them, and redistributing these infections uniformly in the network while keeping the total number of infected nodes fixed (these configurations are named SRI for small and BRI for large communities).

\subsection{Statistics}  
\label{Methods_D}

To analyze the recruitment trees, we measure the total number of participants $\Omega$ (i.e.\ the sample-size), and the size $S_i$ and the number of generations (or waves) $W_i$ of each recruitment tree, starting from a seed node $i$.

The proportion of individuals in the population with a certain feature $A$ ($\hat{P}_A$) is estimated by using the RDSII estimator~\cite{Volz2008}:

\begin{equation}
\label{eq:01}
\hat{P}_A = \frac{\sum\limits_{i \in A \cap N} k_i^{-1}}{\sum\limits_{i \in N} k_i^{-1}}
\end{equation}

where $k_i$ is the reported degree of an individual $i$ in the social network. We thus define:

\begin{equation}
\label{eq:02}
\theta=\sum\limits_{j=1}^m \frac{\hat{P}^{j}_A}{m}
\end{equation}

as the average estimate of the prevalence of $A$ for $m$ simulations with the same set of parameters, with standard deviation given by $\sigma$. Complementary, we define the average bias $\delta$, i.e.\ the difference between the estimate of the prevalence of $A$ and the true prevalence of $A$, for $m$ simulations, as:

\begin{equation}
\label{eq:03}
\delta = \sum\limits_{j=1}^m \frac{|\hat{P}^{j}_A - P_A |}{m}
\end{equation}

In the results, we show the relative bias in respect to the true value of the prevalence, i.e.\ we show $\Delta=\delta/0.25$. The design effect~\cite{Lohr2009} $D.E.$ is defined as:

\begin{equation}
\label{eq:04}
D.E.= \frac{Var(\hat{P}_A)_{\text{RDS}}}{Var(\hat{P}_A)_{\text{SRS}}}
\end{equation}

where $Var(\hat{P}_A)_{\text{RDS}}$ is the variance of the estimator $\hat{P}_A$ using RDS, and $Var(\hat{P}_A)_{\text{SRS}}$ is the variance of the same estimator $\hat{P}_A$ using simple uniform sampling (SRS), i.e.\ the same number of nodes (as in the RDS sample) is uniformly selected in the study population. The design effect thus measures the number of the sample cases necessary to obtain the same statistics as if a simple random sample was used. In our study, $m=500$ ($50$ RDS simulations for each of the $10$ generated network with fixed parameters, and $500$ RDS simulations for each of the empirical networks).

\section{\large Results}

\begin{figure*}[htb]
\centering
\includegraphics[scale=1.1]{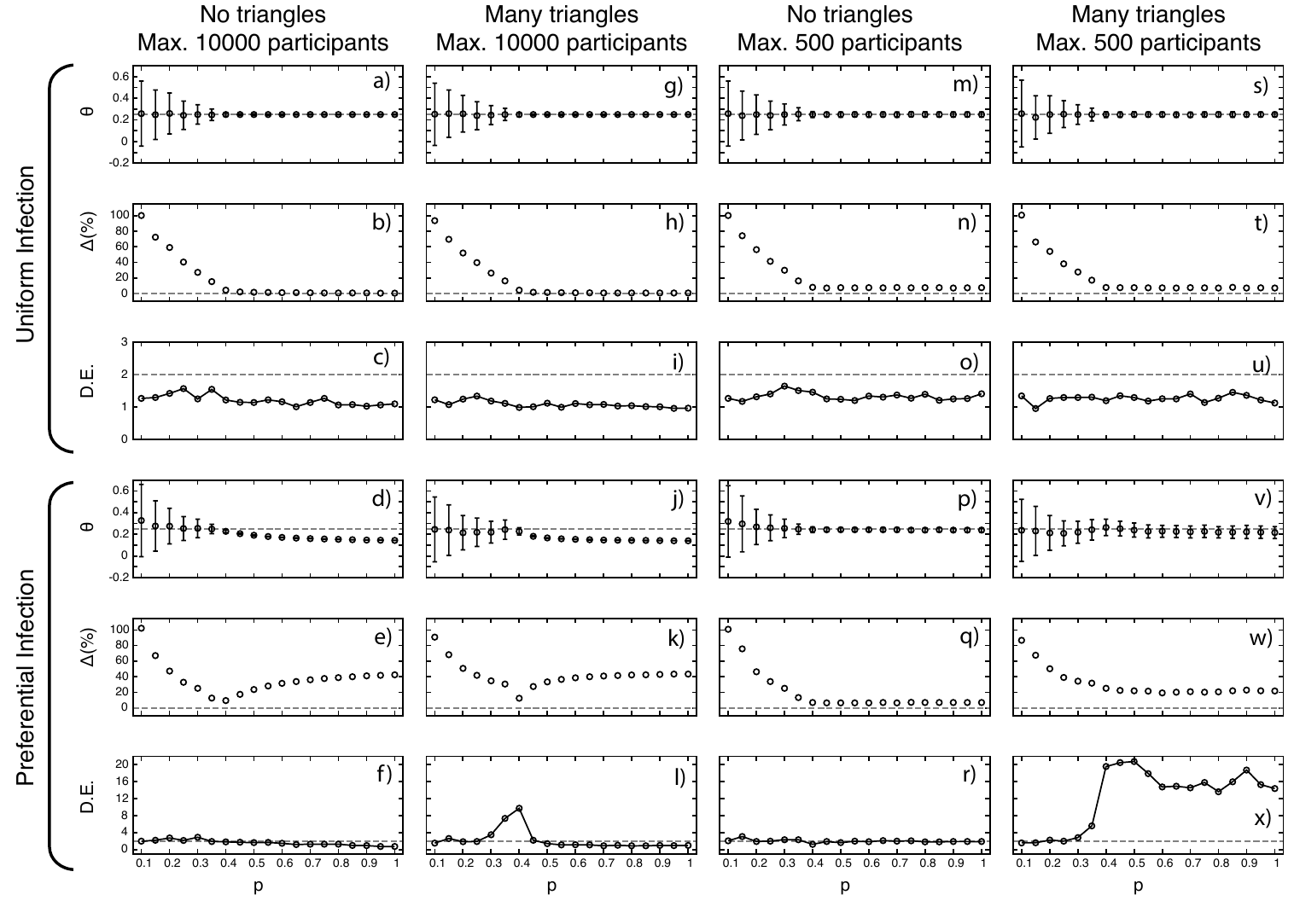}
\caption{\textbf{RDS estimates for networks with triangles.} The panel shows a,d,g,j,m,p,s,v) the RDS estimator $\theta$ (Eq.~\ref{eq:02}) and the respective standard deviation $\sigma$, b,e,h,k,n,q,t,w) the average bias $\Delta$ (Eq.~\ref{eq:03}), and c,f,i,l,o,r,u,x) the design effect $D.E.$ (Eq.~\ref{eq:04}) in respect to the response-rate $p$. In the 1st and 3rd columns, the underlying networks have no triangles and recruitment is limited respectively to $10000$ and to $500$ participants. In the 2nd and 4th columns, the networks have a large number of triangles and recruitment is limited respectively to $10000$ and to $500$ participants (See Section~\ref{Methods_A}). In all cases, $25\%$ of the population is infected with $A$, either following the protocol RI, i.e.\ infections are uniformly spread (top 3 rows), or protocol PRI, i.e.\ infections occur preferentially in high degree nodes (bottom 3 rows) (See Section~\ref{Methods_C}). Dotted horizontal lines are eye-guides. }
\label{fig:03}
\end{figure*}

We first discuss the statistics of recruitment trees for synthetic networks with various levels of clustering and community structure. We then analyze the performance of the RDSII estimator for different network structures and for different scenarios of prevalence of the infection $A$. This analysis is followed by results on the convergence of the estimator for increasing sample size on networks with strong community structure. Afterwards, we study the RDS performance considering the same scenarios of prevalence of the infection using real social networks and conclude the results section showing the increased bias as a consequence of running a single recruitment tree per time.

\subsection{Recruitment trees}  
\label{Results_A}

We first look at some statistics of the recruitment trees in the case that the entire target population can potentially be recruited, i.e.\ the recruitment only stops if no new subject is recruited or if the network is exhausted (everyone is recruited). Since the population is fixed to $10000$ individuals, this limiting case provides us the maximum possible coverage of the sampling for a given configuration of the RDS. In the reference case (Fig.~\ref{fig:02}a-e), only the degree distribution is fixed and the nodes are uniformly connected (configuration model, see Section~\ref{Methods_A}). In this case, if every recruited individual responds to the survey, i.e.\ $p=1.0$ (see Section~\ref{Methods_B}), nearly all the population is recruited. The recruitment dynamics however is not robust to variations in the response-rate, for example, in our simulations, for $p=0.7$, only about $80\%$ of the population is recruited, and this percentage falls to negligible values if $p<0.4$~\cite{Newman2002, Malmros2015}. Successful recruitment in fact occurs only if $p\gtrsim 0.35$ in the absence of any (or negligible) triangles and community structure. We observe a broad distribution in the size of the recruitment trees (Fig.~\ref{fig:02}b,c). There is a relatively high chance for the recruitment trees to break down quickly and thus to contain only a few individuals. This typically happens when a recruitment tree reaches a high-degree node. High-degree nodes are easily reachable because they have many connections. Once the first recruitment tree passes through one of these high-degree nodes, they become unavailable. Consequently, the recruitment trees arriving afterwards simply die out as soon as they reach these nodes. At the same time, a few recruitment chains persist long enough and generate large trees, potentially sampling large parts of the network from a single initial seed. As expected, there is a characteristic peak in the number of waves (Fig.~\ref{fig:02}d,e).

\begin{figure}[htb]
\centering
\includegraphics[scale=1.1]{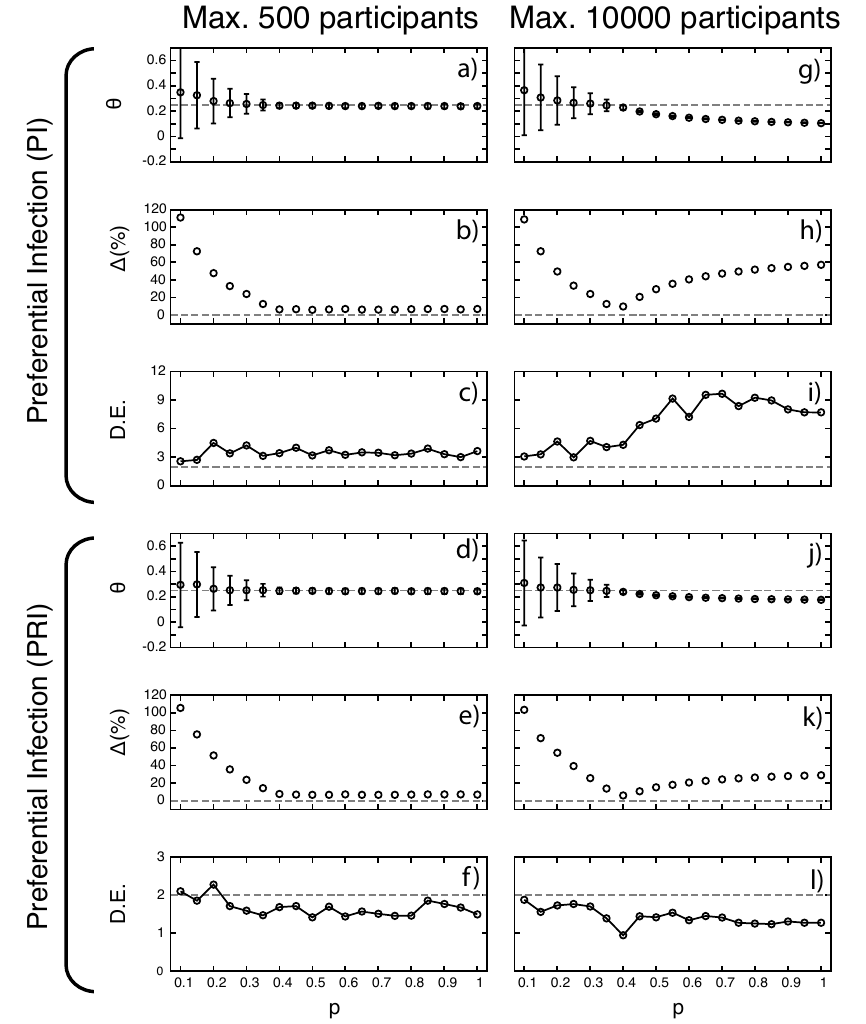}
\caption{\textbf{RDS estimates for networks with weak community structure.} The panel shows a,d,g,j) the RDS estimator $\theta$ (Eq.~\ref{eq:02}) and the respective standard deviation $\sigma$, b,e,h,k) the average bias $\Delta$ (Eq.~\ref{eq:03}), and c,f,i,l) the design effect $D.E.$ (Eq.~\ref{eq:04}). The underlying networks have a few number of triangles and weak community structure (See Section~\ref{Methods_A}), and recruitment is limited respectively to $500$ (1st column) and to $10000$ (2nd column) participants. In both cases, $25\%$ of the population is infected with $A$ preferentially towards high degree nodes following either protocol PI (top 3 rows) or protocol PIR (bottom 3 rows) (See Section~\ref{Methods_C}).}
\label{fig:04}
\end{figure}

The increasing level of clustering has some effect in the statistics of the recruitment trees. In particular, in the absence of communities, a large number of triangles improve recruitment for intermediate values of response-rates (Fig.~\ref{fig:02}f-j). Triangles create redundant paths eliminating bottlenecks in the network, as for example, bottlenecks due to high degree nodes. High degree nodes make a large number of contacts and thus connect different parts of the network. As mentioned before, as soon as these nodes are recruited, the recruitment chain may not be able to expand beyond them. On the other hand, if the network has weak (Fig.~\ref{fig:02}k) or strong (Fig.~\ref{fig:02}p,u) community structure, the number of triangles becomes irrelevant, and the level of community structure defines the sample size. In case of strong community structure (with low or large number of triangles), a maximum of $\sim85\%$ of the population may be recruited (Fig.~\ref{fig:02}p,u). Bottlenecks in this case correspond to nodes bridging communities. These bottlenecks cannot be removed by adding triangles, that only produce local network redundancy, but by connecting more nodes between different communities, i.e.\ weaken the community structure. Moreover, strong communities imply that response-rates should be higher (in comparison to the absence of or to weaker communities) for the recruitment chains to take off and gather sufficient participants. If response-rates are bellow $p\sim0.45$, recruitment is insufficient. This is a fundamental issue in realistic settings, meaning that highly clustered (or in other words, highly segregated and marginalized) populations need a bit higher compensation in order to achieve the same sampling size as one would obtain if studying less segregated groups.

We see that irrespective of the number of triangles or level of community structure, lower response-rates cause a relatively larger number of small recruitment trees together with a few waves (Fig.~\ref{fig:02}b,d,g,i,l,n,q,s,v,x). This is not only undesirable because the final sample remains small but also because a few waves is not sufficient for the stochastic process to forget the initial conditions and thus reach the stationary state, the condition in which the estimator is expected to be unbiased. In case of strong community structure (Fig.~\ref{fig:02}s,t,x,y), we note a broader variance in the number of waves suggesting that each seed samples the network non-homogeneously. This may be related to the fact that the communities have different sizes (or number of nodes) and thus the bottlenecks between communities are reached at different times by different recruitment chains.

\subsection{RDS estimates and structure-induced bias}  
\label{Results_B}

\begin{figure*}[htb]
\centering
\includegraphics[scale=1.1]{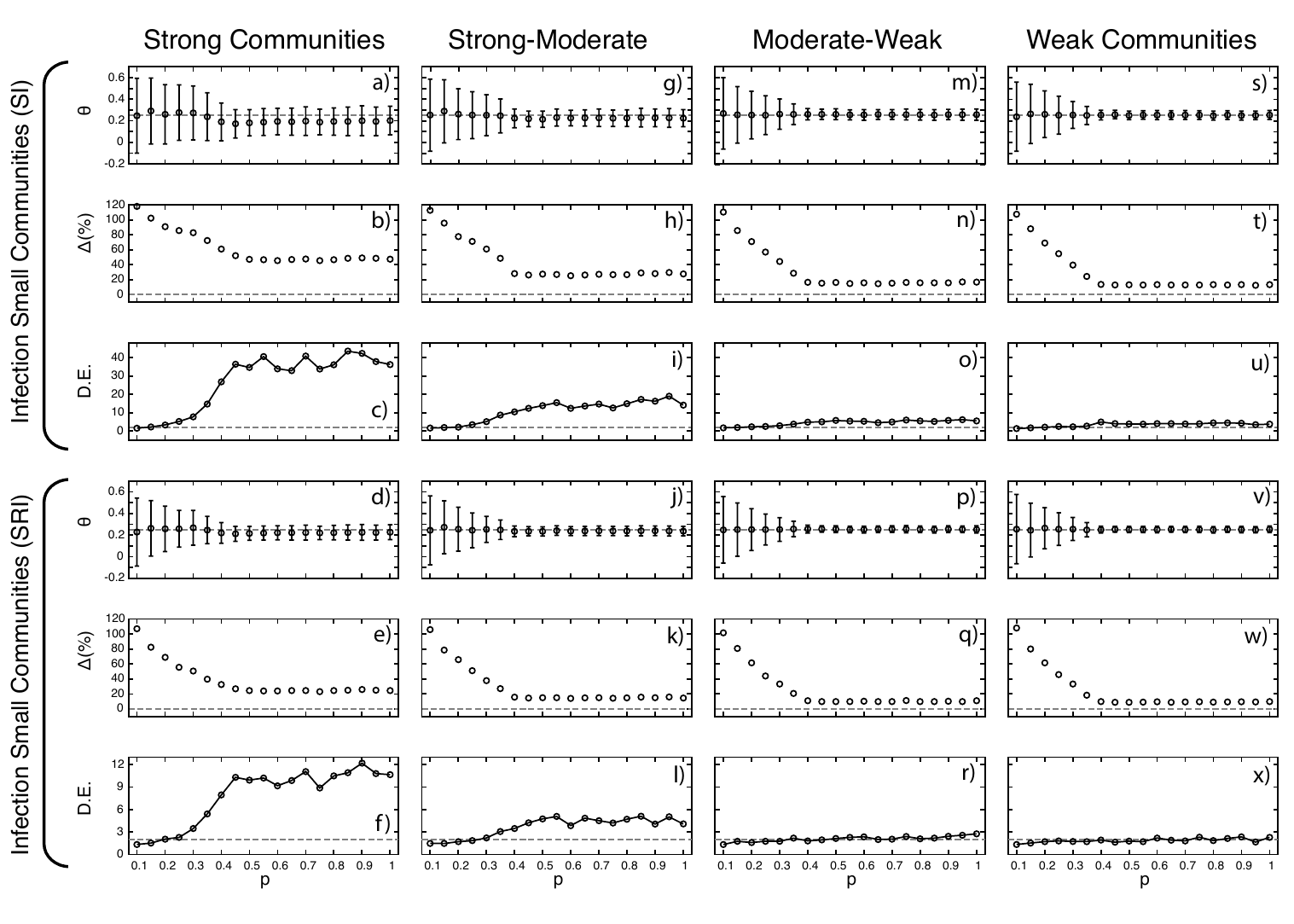}
\caption{\textbf{Prevalence of $A$ in the smallest communities.} The panel shows a,d,g,j,m,p,s,v) the RDS estimator $\theta$ (Eq.~\ref{eq:02}) and the respective standard deviation $\sigma$, b,e,h,k,n,q,t,w) the average bias $\Delta$ (Eq.~\ref{eq:03}), and c,f,i,l,o,r,u,x) the design effect $D.E.$ (Eq.~\ref{eq:02}). The underlying contact networks have various levels of community structure (See Section~\ref{Methods_A}), and recruitment is limited to $500$ participants. In all cases, $25\%$ of the population is infected with a quantity $A$, following either protocol SI (top 3 rows) or protocol SRI (bottom 3 rows) (See Section~\ref{Methods_C}).}
\label{fig:05}
\end{figure*}

To study the impact of the network structure and response-rates in the RDS estimator, we measure four statistics: (i) the average RDSII estimator $\theta$ (Eq.~\ref{eq:02}) and its respective (ii) standard deviation $\sigma$, (iii) the average bias $\Delta$ (see Eq.~\ref{eq:03}), and (iv) the design effect $D.E.$ (Eq.~\ref{eq:04}) (See Section~\ref{Methods_D}). Figures~\ref{fig:03}a-f,m-r show the reference case, i.e.\ the configuration model where the only structure is in the degree sequence and the rest is random (See Section~\ref{Methods_A}). In this reference case, if there is no restriction to the sample size in respect to the size of the target population (i.e.\ up to 10000 individuals may be sampled, but actual sample size depends on the response-rate), the estimator $\theta$ performs well, although with substantial standard deviation $\sigma$ and biases $\Delta$ for $p<0.4$ even if the quantity $A$ is uniformly spread in the network (Fig.~\ref{fig:03}a-c). This is a result of the insufficient sample size for low response-rates.

Individuals with a large number of contacts are believed to be more central in a network~\cite{Newman2010}. These individuals may be for example more likely to get an infection or propagate a piece of information. We thus test an hypothetical scenario where $A$ is concentrated in high-degree nodes (See protocol PRI in Section~\ref{Methods_C}). Note that this assumption may be however completely irrelevant in some contexts, but it is useful to understand the mechanisms of sampling. In this case, the accuracy of the estimator $\theta$ is poor for $p>0.3$, i.e.\ $A$ is under-estimated for both situations, with and without many triangles, and precision is worse for response-rates $p<0.35$ (respectively Fig.~\ref{fig:03}j-l and Fig.~\ref{fig:03}d-f). As before, the poor accuracy is a result of the RDS not recruiting sufficient participants. The under-estimation of the prevalence however suggests that low-degree nodes are not being sufficiently sampled as the sample size gets close to the network size. A substantial bias, given by $\Delta$, is also observed. The design effect varies between 1 and 2, with some exceptions for $p\sim 0.4$ in case of many triangles. If the number of participants is limited to only $500$ individuals, i.e.\ $5\%$ of the total population (a small fraction of the target population, as usually recommended to guarantee unbiased estimates~\cite{Gile2010}), the performance of the estimator $\theta$ and the average bias $\Delta$ improves substantially. However, $A$ remains slightly underestimated and the standard deviation $\sigma$ increases in the case of many triangles irrespective of the response-rates (Fig.~\ref{fig:03}v-x). The cost of this improvement however is a much higher design effect (Fig.~\ref{fig:03}x).

\begin{figure*}[htb]
\centering
\includegraphics[scale=1.1]{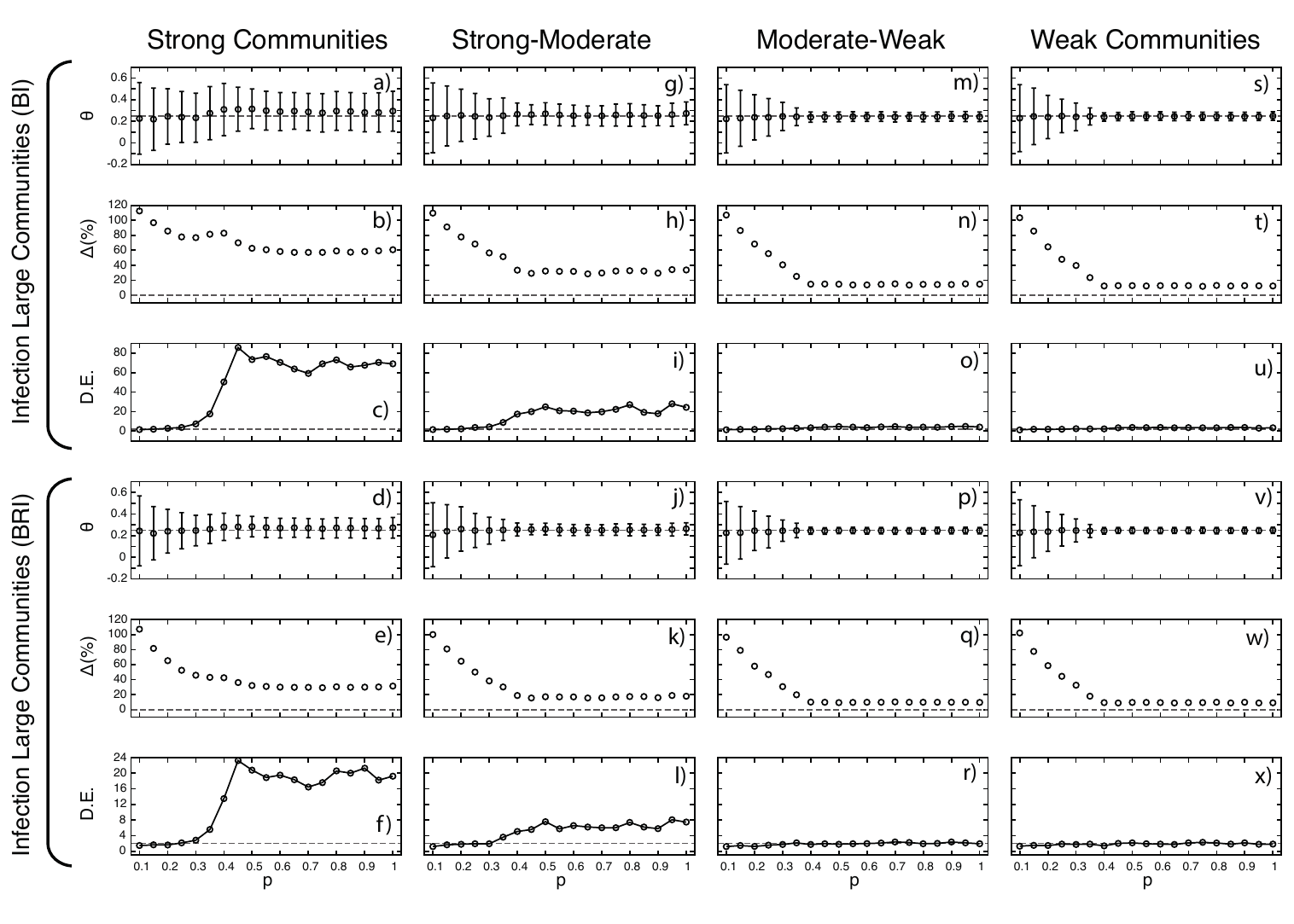}
\caption{\textbf{Prevalence of $A$ in the largest communities.} The panel shows a,d,g,j,m,p,s,v) the RDS estimator $\theta$ (Eq.~\ref{eq:02}) and the respective standard deviation $\sigma$, b,e,h,k,n,q,t,w) the average bias $\Delta$ (Eq.~\ref{eq:03}), and c,f,i,l,o,r,u,x) the design effect $D.E.$ (Eq.~\ref{eq:02}). The underlying contact networks have various levels of community structure (See Section~\ref{Methods_A}), and recruitment is limited to $500$ participants. In all cases, $25\%$ of the population is infected with $A$, following either protocol BI (top 3 rows) or protocol BRI (bottom 3 rows) (See Section~\ref{Methods_C}).}
\label{fig:06}
\end{figure*}

Figure~\ref{fig:04}a-f shows that in networks with weak community structure, if $A$ is concentrated at the high-degree nodes, the estimates remain good for $p\gtrsim 0.2$ if the maximum number of participants is low (up to 500) compared to the total size of the target population. In the limiting case where all individuals can potentially participate (up to 10000), $A$ is slightly overestimated and substantially underestimated respectively for small and large response-rates, being accurate only for moderate values, i.e.\ $p \sim 0.4$ (Fig.~\ref{fig:04}g-l). The results suggest that for larger response-rates, there is a significant under-representation of low-degree nodes in the final sample. This happens because low-degree nodes become increasingly more difficult to sample as the sample size gets close to the network size (causing finite-size effects). Biases are also larger if the community structure is stronger because the recruitment chains die out before exploring some of the communities. Altogether, these results are in accordance with previous recommendations that the sample size should be much smaller than the size of the target population~\cite{Gile2010} in order to achieve good estimates using the RDSII estimator. Some caution however should be pointed out since it is not straightforward to know in advance the size of the target population and thus to estimate the optimal sample size in respect to the target population. If too many subjects are recruited, relatively to the size of the target population, saturation occurs and the network structure induces biases in the estimator due to finite-size effects.

We now simulate scenarios where the variable $A$ is concentrated in specific communities, irrespective of the degree of the nodes. This is a reasonable assumption considering that an infection (or other particular quantities) may affect only the population of some geographical region, or for example, a particular group of injecting drug users among MSM may be sharing contaminated paraphernalia. By using the know structure of each network, we select $25\%$ of the nodes associated to the smallest communities and infect them with the quantity $A$ (See Section~\ref{Methods_C}). In this setting, the prevalence is underestimate and the estimator has relatively large deviations (Fig.~\ref{fig:05}a,g) for strong and strong-moderate community structure. Estimators improve for weaker community structure (Fig.~\ref{fig:05}m,s). Even for weak community structure, the minimum average bias is about $15\%$ (Fig.~\ref{fig:05}t), being at least $45\%$ in case of strong communities (Fig.~\ref{fig:05}b) for $p=1.0$. For lower response-rates, the bias gets substantially larger, as in the previous experiments. The design effect is also significantly affected for any level of community structure (Fig.~\ref{fig:05}c,i,o,u). This means that for strong communities, for example, in order to have the same statistics as if a standard simple random sample was performed, the RDS needs up to 40 times the same sample size. Furthermore, if we redistribute the infection of $40\%$ randomly chosen infected nodes to decrease homophily, the overall quality of the statistics improves but still with significant bias, and larger standard deviation and design effect for stronger community structure (Fig.~\ref{fig:05}d-f,j-l,p-r,v-x).

\begin{figure}[htb]
\centering
\includegraphics[scale=0.85]{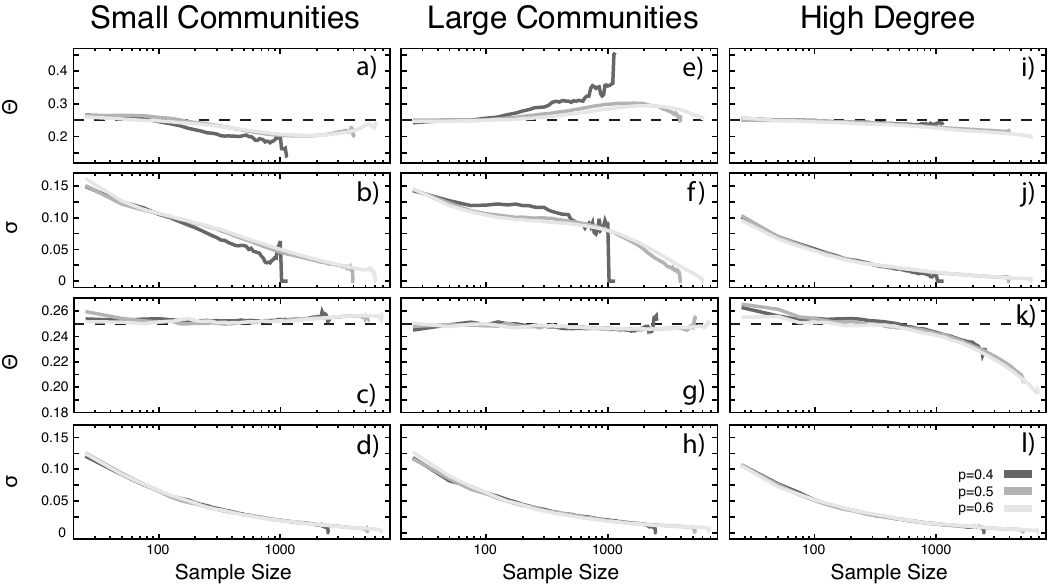}
\caption{\textbf{Estimates of prevalence and sample size.} The panel shows the estimator $\theta$ and the respective standard deviation $\sigma$ for networks with a,b,e,f,i,j) strong and c,d,g,h,k,l) weak community structure. In the 1st column, $A$ is concentrated in the small communities (SRI protocol), in the 2nd column, $A$ is concentrated in the large communities (BRI protocol), and in the 3rd column, $A$ is concentrated in the high degree nodes (PRI protocol).}
\label{fig:07}
\end{figure}

\begin{figure}[htb]
\centering
\includegraphics[scale=1.1]{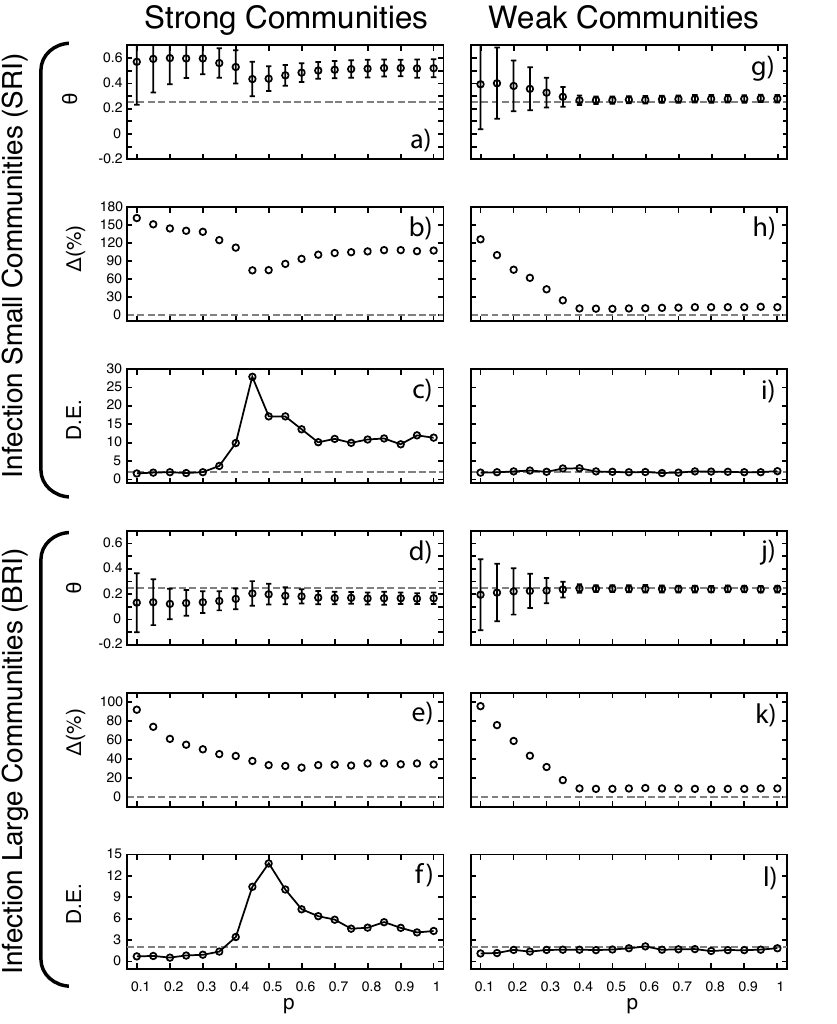}
\caption{\textbf{RDS estimates with seeds selected inside small communities.} The panel shows the RDS estimator $\theta$ (Eq.~\ref{eq:02}) and the respective standard deviation $\sigma$, the average bias $\Delta$ (Eq.~\ref{eq:03}), and the design effect $D.E.$ (Eq.~\ref{eq:04}). In all cases, $25\%$ of the population is infected with $A$ in a-c,g-i) the smallest communities and in d-f,j-l) the largest communities (See Section~\ref{Methods_C}). a-f) networks with strong communities and g-l) networks with weak communities.}
\label{fig:08}
\end{figure}

\begin{figure}[htb]
\centering
\includegraphics[scale=1.1]{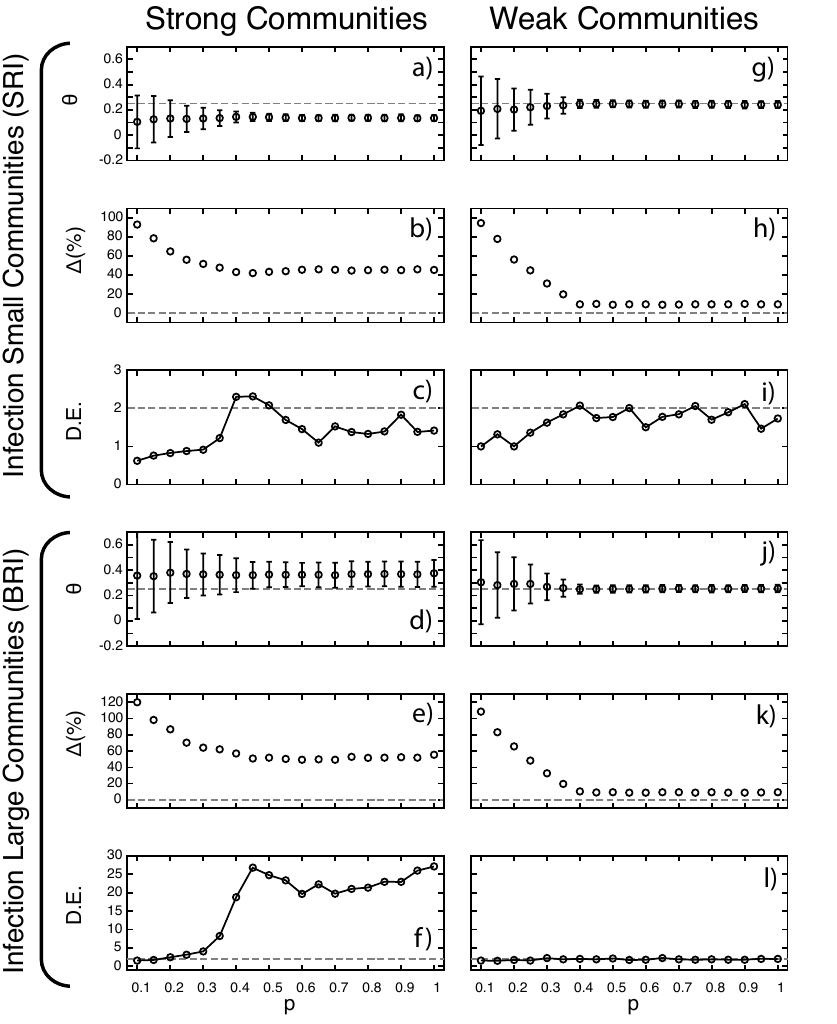}
\caption{\textbf{RDS estimates with seeds selected inside large communities.} The panel shows the RDS estimator $\theta$ (Eq.~\ref{eq:02}) and the respective standard deviation $\sigma$, the average bias $\Delta$ (Eq.~\ref{eq:03}), and the design effect $D.E.$ (Eq.~\ref{eq:04}). In all cases, $25\%$ of the population is infected with $A$ in a-c,g-i) the smallest communities and in d-f,j-l) the largest communities (See Section~\ref{Methods_C}). a-f) networks with strong communities and g-l) networks with weak communities. }
\label{fig:09}
\end{figure}

On the other hand, we can assume that $A$ is unlike to occur in small communities because, for example, nodes associated to these communities are simply less likely to get an infection due to isolation. Social control is also often higher in small groups. It may therefore be easier to behave in certain ways in larger groups. People who want to or who have particular behaviors or traits may thus decide to move to larger groups. To simulate this hypothetical scenario, we infect $25\%$ of the nodes in the largest communities (See Section~\ref{Methods_C}). Figure~\ref{fig:06}a shows that $A$ is overestimated for $p>0.3$ for strong community structure. These estimates improve for weaker communities, also resulting on smaller standard deviations (Fig.~\ref{fig:06}g,m,s) for larger response-rates. The standard deviation is generally slightly larger in this case in comparison to the case where $A$ is concentrated in the small communities. The design effect is very high for strong community structure (Fig.~\ref{fig:06}c,i), even if homophily is reduced (Fig.~\ref{fig:06}f,l).

We perform the same analysis using networks with the same configuration studied until now but with higher clustering coefficient (between 0.5 and 0.6) and the results remain quantitatively the same (apart for a few fluctuations). This finding reinforces the previous observation that triangles have a relatively small impact in RDS if communities are present in the network. Altogether, these results show the key difference between clustering and homophily that was mentioned in the Introduction. In both scenarios, the network community structure and the number of triangles are the same, and homophily is high. In the later case (Fig.~\ref{fig:06}) homophily occurs inside the largest communities whereas in the first case (Fig.~\ref{fig:05}) it happens in the smallest communities. The structure-induced biases however remain relatively high even if the homophily is reduced by redistributing a fraction of the infections.

\subsection{Convergence and sample size}
\label{Results_C}

In the previous section, we have studied the bias induced by the network structure and the response-rates. If we fix the response-rate, each realization of the simulation generates a different sample size due to the stochastic nature of the process. In this section, therefore, we fix the response-rate and analyze the effect of the sample size on the estimator. Since recruitment may stop at different times on each simulation, here we estimate the mean and standard deviation for sample size $S$ using only simulations in which the recruitment reaches this size $S$. This means that the estimates for large sample sizes have less data points (to calculate the mean) than those for small sample sizes. Previous studies report that in real settings, response-rates may vary between $0.3$ (for female sex-workers) and $0.7$ (for MSM), with mean and median at about $0.5$~\cite{Gile2015}. We thus study 3 scenarios for the response-rates: $p={0.4,0.5,0.6}$. Figure~\ref{fig:07} shows that for strong community structure, $\theta$ is slightly overestimated for sample sizes smaller than $100$ and underestimated for larger sample sizes if $A$ is concentrated in the smallest communities (Fig.~\ref{fig:07}a,b). On the other hand, the prevalence of $A$ is overestimated for sample sizes larger than $100$ if $A$ is concentrated in the largest communities (Fig.~\ref{fig:07}e,f). In both cases, the mismatch is maximized when the sample size is between about $10\%$ and $30\%$ (i.e.\ 100 and 1500 participants respectively) of the study population. If $A$ is concentrated in the high degree nodes, $\theta$ is underestimated for increasing sample size (Fig.~\ref{fig:07}i,j) but not as much as for the previous cases. On the other hand, if the community structure is weak, the estimator performs well (with slight over- and under-estimation of the prevalence for small (Fig.~\ref{fig:07}c,d) and large (Fig.~\ref{fig:07}g,h) communities, except in the case of $A$ being concentrated in high degree node (Fig.~\ref{fig:07}k,l). In this case, the estimates are only good in the range of sample sizes between $100$ and $1000$ nodes.

\begin{figure*}[htb]
\centering
\includegraphics[scale=0.6]{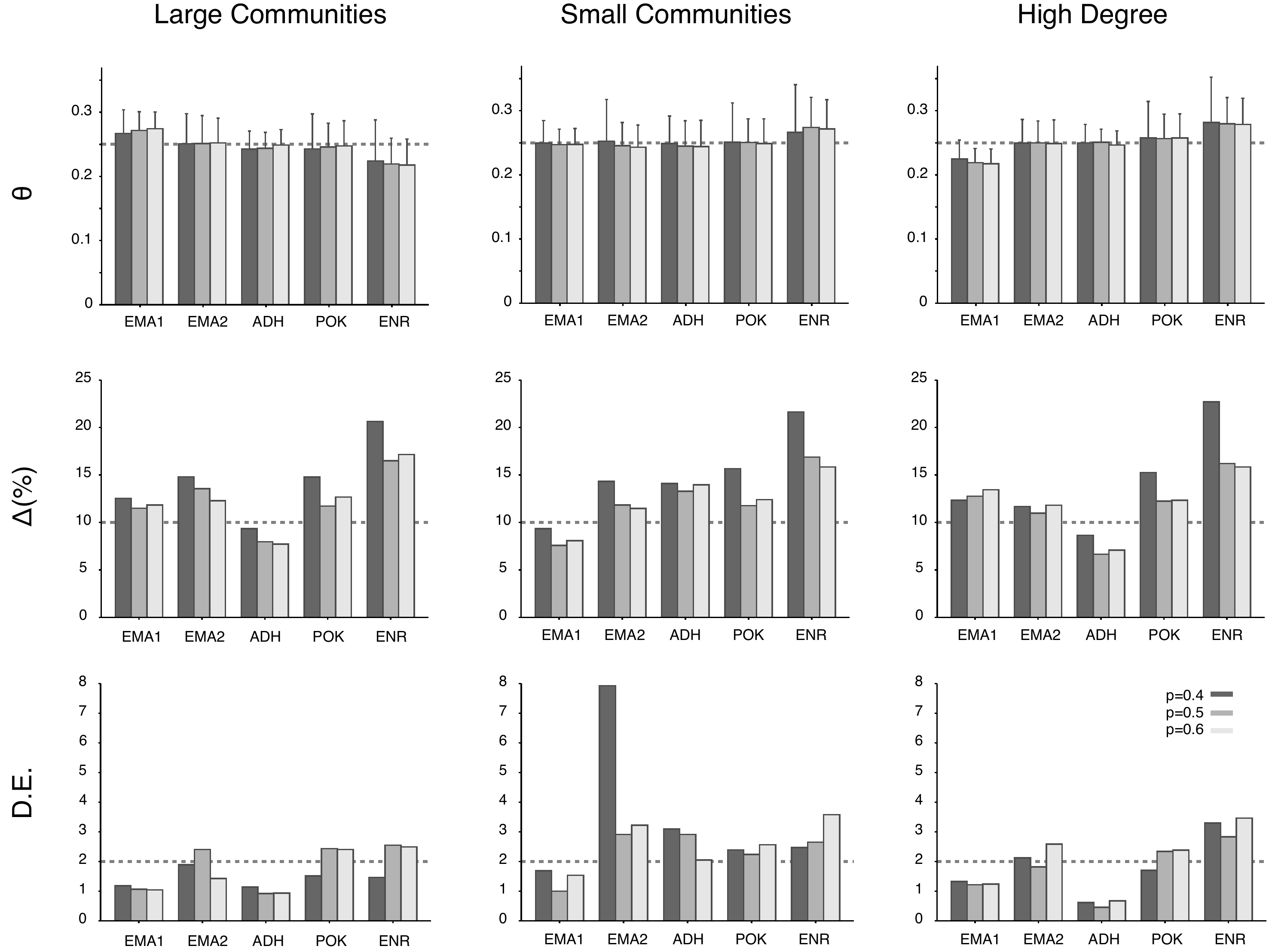}
\caption{\textbf{RDS estimates for empirical networks.} The panel shows the RDS estimator $\theta$ (Eq.~\ref{eq:02}) and the respective standard deviation $\sigma$, the average bias $\Delta$ (Eq.~\ref{eq:03}), and the design effect $D.E.$ (Eq.~\ref{eq:04}). The contact networks are gathered empirically and correspond to different types of social relation and population size (See Section~\ref{Methods_A}). Recruitment is limited to $500$ participants and the response-rate $p$ covers realistic values. In all cases, $25\%$ of the population is infected with a quantity $A$, either in the largest communities (1st column), or smallest communities (2nd column), or the high degree nodes (3rd column) (See Section~\ref{Methods_C}).}
\label{fig:10}
\end{figure*}

\subsection{Seed-induced bias}
\label{Results_D}

We have assumed so far that seeds are uniformly chosen within the target population. While this is a reasonable standard assumption in theoretical studies, it is hardly met in real contexts because the inherent fact that the study population is hard-to-reach and seed selection is non trivial~\cite{Wylie2013}. A biased selection of seeds can increase the bias in the RDS estimators as shown in Figs.~\ref{fig:08} and~\ref{fig:09}. If seeds are selected only between subjects associated to small communities (here defined as communities with less than 200 members), recruitment chains are generally unable to reach beyond those communities and thus the prevalence is overestimated when the infection is concentrated in the smaller communities (Fig.~\ref{fig:08}a-c). On the other hand, the prevalence is underestimated if the infection is concentrated in the larger communities (Fig.~\ref{fig:08}d-f). The mismatch in the estimators are particularly significant if the community structure is stronger, however, the prevalence is also strongly biased for low response-rates (and weakly biased for high response-rates) even if the community structure is weak (Fig.~\ref{fig:08}g-l). This is in contrast to the our previous findings when seeds are uniformly sampled (Fig.~\ref{fig:05} and~\ref{fig:06}).

If one selects the seeds in the largest communities (here defined as communities with more than 500 members), recruitment chains tend to stay within the largest communities, which leads to an under-estimation of the prevalence and relatively high biases if the infection is mostly prevalent in the small communities (Fig.~\ref{fig:09}a-c). The prevalence is overestimated, however, if the infection is mostly prevalent in the largest communities (Fig.~\ref{fig:09}d-f). Results improve for weak community structure, but also in this case, biases and large standard deviations are observed for low response-rates (Fig.~\ref{fig:09}g-l). Note that in these experiments homophily is relatively weak since we use protocols SRI and BRI.

\subsection{Empirical networks}
\label{Results_E}

In the previous sections, we have studied the impact of various levels of community structure and number of triangles in RDS estimates in contact networks generated using theoretical models. Although the algorithms used to generate the synthetic networks include several properties of real-life networks, empirical networks, with their own sampling and scope limitations, contain correlations that may be challenging to reproduce theoretically. In this section, we analyze the RDS performance using real-life human contact networks in order to be able to extend the conclusions to real scenarios. Following the same protocols to infect preferentially the largest (BRI protocol) or the smallest (SRI protocol) communities, or the high degree nodes (PRI protocol), we find that in most studied networks, RDS performs well to estimate the mean prevalence in these hypothetical scenarios (although the standard deviations are relatively large), with a small variation for different response-rates (Fig.~\ref{fig:10}). The estimates are worse for EMA1 and ENR datasets, respectively, the smallest and the largest networks (See Table~\ref{tab:01} in Section~\ref{Methods_A}). We see that the average bias is larger than $10\%$ with a few exceptions. It is also typically larger for $p=0.4$. The design effect is generally somewhere between 1 and 3 (one exception for $p=0.4$ and EMA2), a result inline to previous suggestions that a design effect of $2$ may be used as a general guideline on unknown populations~\cite{Salganik2006}.

\begin{figure}[htb]
\centering
\includegraphics[scale=1.1]{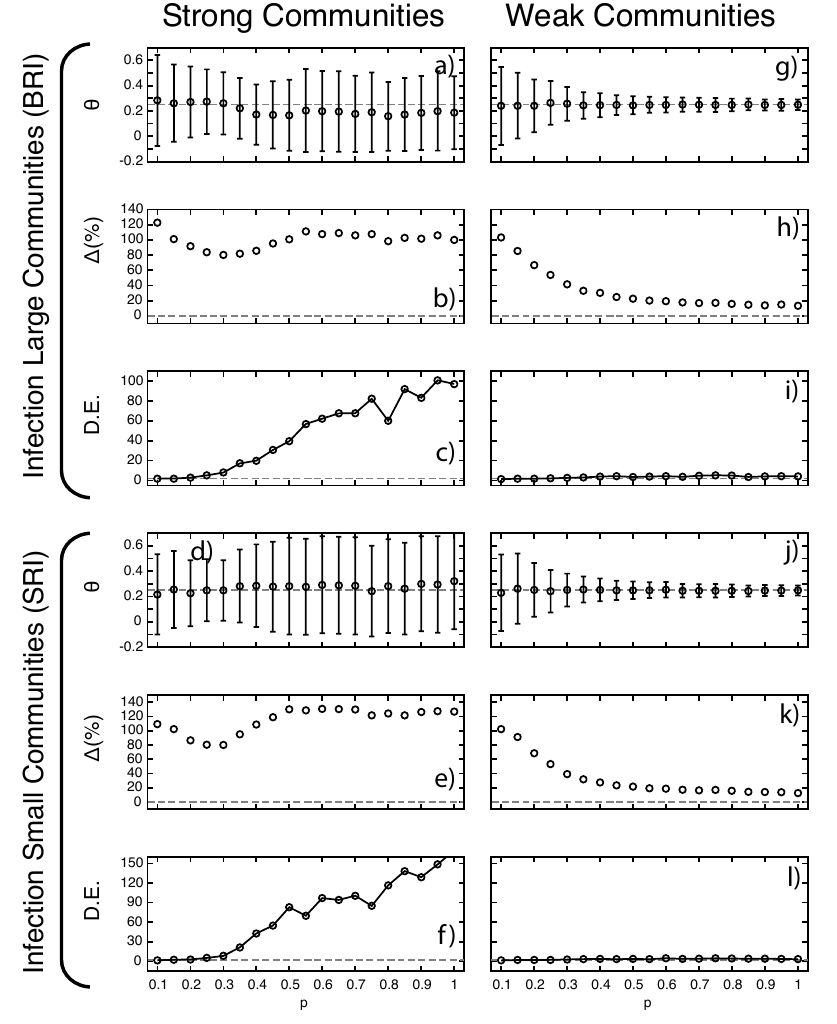}
\caption{\textbf{RDS estimates with improved seed sampling mechanism.} The panel shows the RDS estimator $\theta$ (Eq.~\ref{eq:02}) and the respective standard deviation $\sigma$, the average bias $\Delta$ (Eq.~\ref{eq:03}), and the design effect $D.E.$ (Eq.~\ref{eq:04}). Recruitment is limited to $500$ participants. In all cases, $25\%$ of the population is infected with $A$ in a-c,g-i) the largest communities  and in d-f,j-l) the smallest communities (See Section~\ref{Methods_C}). a-f) networks with strong communities and g-l) networks with weak communities.}
\label{fig:11}
\end{figure}

\subsection{Non-simultaneous seed sampling}
\label{Results_F}

Recruitment trees often break down after a few waves in real settings. As discussed above, this is not only a consequence of low response-rates but also the effect of multiple recruitment trees, originating from different seeds, bumping each other and then dying out. A practical solution to obtain sufficiently large sample sizes is to restart the recruitment with new seeds as soon as the recruitment stops. In this section we test the effect on the estimators if we start a new seed after the previous recruitment chain has completely stopped, i.e.\ seeds start at different points in time. We assume that the recruitment chain stops either naturally or after reaching a certain size. In particular, we test the case when the target sample size is $500$ participants out of the population of $10000$ people, using 10 seeds as done in the previous sections. Each seed is allowed to recruit (successfully) at maximum 50 participants, and a new seed is only selected (uniformly among non-recruited nodes) when the current recruitment stops. As usual, the same person may participate only once.

Figure~\ref{fig:11} shows that the average estimator is affected and the prevalence is under-estimated if $A$ is concentrated in the largest communities and over-estimated if $A$ is more likely in smaller communities. The standard deviations are relatively large in case of strong communities (Fig.~\ref{fig:11}a,d) and decreases, but still maintaining relatively large values, for weaker communities (Fig.~\ref{fig:11}g,j). The average bias and design effect substantially change in comparison to the case when seeds are selected simultaneously, particularly for strong community structure. The restarting of seeds introduces mixing, or equivalently, random links, in the network structure~\cite{Lambiotte2012}. Moreover, the restriction in the size of the recruitment trees possibly inhibits the sampling process to reach the stationary state, a factor known to cause biases in the estimators~\cite{Volz2008}. The major consequence of these very large biases is that one is not sure that a single RDS experiment (as is usually the case in reality) provides a reliable estimate. Selecting exactly one new seed after the current seed has being exhausted is an hypothetical situation. This extreme case however illustrates that the non-simultaneous selecting of seeds may increase the biases substantially unless only a few re-starts occur. We expect that more realistic scenarios (e.g.\ initially selecting multiple seeds simultaneously and eventually selecting a few new seeds if the original recruitment trees die out) lie somewhere between this case and the simultaneous seed sampling studied in Section~\ref{Results_B}.

\section{\large Discussions}

Respondent-driven sampling has been proposed as an effective methodology to estimate the prevalence of variables of interest in hard-to-reach populations. The approach exploits information on the social contacts for both recruitment and weighting in order to generate accurate estimates of the prevalence. Social networks however are not random but contain patterns of connectivity that may constrain the cascade of sampling. In particular, nodes have a high heterogeneity in the number of contacts, and networks typically have many triangles and a community structure.

In this paper, we have studied the bias induced by community structure and network triangles in the RDS by using both synthetic and empirical network structures with various levels of clustering, size, degree heterogeneity, and so on. We have also analyzed the impact of various response-rates in the estimators and quantified the relative bias for combinations of parameters. Altogether, we have identified that the structure of social networks have a relevant impact on RDS leading to potential biases in the RDS estimator. The estimator generally performs sufficiently well if response-rates are sufficiently high, the community structure is weak and the prevalence of the variable of interest is not much concentrated in some parts of the network (low homophily). The high heterogeneity of the network communities implies that sampling chains may get constrained to certain parts of the network and thus the prevalence of the infection may be either under- or over-estimated depending on which part of the network concentrates more infections. Some parts of the network may only be accessed through tight bottlenecks, i.e.\ key individuals that bridge the small well-hidden sub-groups and the rest of the population. If these bridging nodes are not willing to participate in the recruitment or once they are recruited, recruitment trees get trapped within a group of nodes, oversampling them, and generating biases.

The structure of empirical networks may vary in different contexts. Consequently, the expected biases may be also lower or higher for certain social networks. In particular, biases should increase for sparser networks because less paths are available between the nodes. In other words, there are more bridging nodes maintaining the network connected and thus the recruitment becomes more sensitive to lower response-rates. Similarly, lower biases are expected in denser networks. The number of network communities and the distribution of community sizes may be also different than the ones we consider. Many small communities have a significant effect in the sampling, increasing the biases, because they imply on the existence of many bridging nodes and higher chances to divert or break down the recruitment. We have also assumed that those people who choose to not participate in the first invitation may be invited again. This possibly introduces a positive correlation between chance to answer the survey and the degree of the node, i.e.\ a tendency to oversample high degree nodes not related to clustering or community structure. Since this is not possible for response-rates $p \neq 1$ and we generally observe relatively similar results for decreasing $p$, if this effect occurs, it is only relevant for low response-rates $p$. This may however explain why we generally observe a transition in the average bias at values above the critical response-rates (the point where a significant number of individuals is recruited). On the other hand, the effect of clustering and communities should be even higher if we assume that people cannot be invited more than once (or equivalently, if someone refuses to participate the first time, it may refuse the following times as well) since this is further blocking the access to certain parts of the network.

To understand the effect of the participation probability $p$, we may consider the simple case where a one single coupon is exchanged between individuals, and the sampling is done with replacement~\cite{Volz2008}. In that case, the stochastic process is equivalent to a random walk process if $p=1$. The probability $p_i$ of finding a coupon with person $i$ is driven by the rate equation 
\begin{equation}
\label{ctrw}
\dot{p}_{i} = \sum_{j} \frac{A_{ij}}{k_j} p_{j} - p_i
\end{equation}
where $A_{ij}$ is the adjacency matrix of the social network. In the case of undirected and unweighted networks, where each link is reciprocated and carries the same importance, the element matrix $(i,j)$ of the matrix is equal to 1 if there is a link between $i$ and $j$ and zero otherwise. The study of this stochastic process has a long tradition in applied mathematics and statistical physics (e.g.~\cite{Klafter2011, Delvenne2010}). Relevant to our results, it is known that the system converges to equilibrium if the underlying network is connected~\cite{Volz2008}. In this regime, nodes would be visited by coupons with a probability proportional to their degree and the whole network is explored, independently on the initial conditions. Equilibrium is reached after a characteristic time scale $\tau$ defined as $1/\lambda_2$, where $\lambda_2$ is the first non-zero eigenvalue of the Laplacian matrix driving $p$ in Eq.~(\ref{ctrw}). This time scale is associated to the presence of a bottleneck (the bridging nodes) between two strongly connected communities in the network. For times smaller than $\tau$, the random walk has essentially explored almost uniformly one single community, but has not sufficiently explored the other one. This time scale therefore provides us with a way to estimate the minimal value of $p$ needed for the whole graph to be sampled, that is $1-p < \lambda_2$. The case of sampling with restart is related to the process of random walk with teleportation. In that case, the choice of the seed where to restart the process is known to affect the statistical properties of the sampling of the network~\cite{Lambiotte2012}. A future theoretical exercise is to adapt those ideas to this context in order to improve the RDS estimators on situations where restarting is necessary. Furthermore, using non-backtracking random walks may be a possible theoretical direction to model RDS considering sampling without replacement. Those random walks avoid to go back from where they come from, at the previous step, and they are known to explore the network faster~\cite{Alon2007}.

Finally, the results of our numerical exercise suggest some general recommendations for studies in real settings: i. Experimental researchers should be aware of the potential critical bridge nodes in the study population, which may vary according to the characteristics of the population; ii. Experimental researchers should aim to response-rates at least above 0.4 in order to reduce the associated biases and uncertainty of the estimates. This recommended response-rate may be increased if more coupons are used; iii. Attention should be taken on selecting the seeds as uniformly as possible, particularly aiming to avoid many seeds either in the small or in the large groups (typically the most reachable individuals). The temptation to start all seeds within well-hidden groups may cause the recruitment to not move beyond these groups; iv. Restarting the seeds (to get larger sample sizes) during the ongoing recruitment should be generally avoided. A better strategy may be to either start the experiment with more seeds or to increase response-rates to avoid dropouts.

\section*{\large Acknowledgements} LECR is a FNRS charg\'e de recherches. LECR and AET thank VR for financial support.

\section*{\large Author Contributions} LECR and FL conceived the study. LECR performed the simulations and analyzed the results. LECR, AET, RL, FL wrote the manuscript. 

\section*{\large Additional information} The authors declare that they have no competing financial interests. Correspondence and requests for materials should be addressed to LECR~\mbox{(luis.rocha@ki.se)}.


\begin{thebibliography}{10}
\expandafter\ifx\csname url\endcsname\relax
  \def\url#1{\texttt{#1}}\fi
\expandafter\ifx\csname urlprefix\endcsname\relax\def\urlprefix{URL }\fi
\expandafter\ifx\csname href\endcsname\relax
  \def\href#1#2{#2} \def\path#1{#1}\fi

\bibitem{Sudman1988}
S. Sudman, M. G. Sirken, C.D. Cowan Sampling rare and elusive populations Science 240 991-996 (1988)

\bibitem{Heckathorn1997}
D. D. Heckathorn Respondent-Driven Sampling: {A} New approach to the study of hidden populations Soc. Prob. 44(2) 174-199 (1997)

\bibitem{Volz2008}
E. Volz and D. D. Heckathorn Probability based estimation theory for respondent driven sampling J. Off. Statist. 24 79-97 (2008)

\bibitem{Semaan2010}
S. Semaan Time-space sampling and respondent-driven sampling with hard-to-reach Populations Methodological Innovations Online 5(2) 60-75 (2010)

\bibitem{Abdul2006} 
A. Abdul-Quader, D. Heckathorn, C. McKnight, H. Bramson, C. Nemeth, K. Sabin, K. Gallagher, D. Des Jarlais Effectiveness of respondent-driven sampling for recruiting drug users in new york city: Findings from a pilot study. J. Urban Health 83 (3) 459-476 (2006)

\bibitem{Salganik2006}
M. J. Salganik Variance estimation, design effects, and sample size calculations for respondent-driven sampling. J. Urban Health: Bull. N. York Acad. Med. 83 (7) i98-i110 (2006)

\bibitem{Gile2010}
K. J. Gile and M. S. Handcock Respondent-driven sampling: An assessment of current methodology. Sociol. Methodol. 40 285-327 (2010)

\bibitem{Lu2012}
X. Lu, L. Bengtsson, T. Britton, M. Camitz, B. J. Kim, A. Thorson, F. Liljeros The sensitivity of respondent-driven sampling J. R. Statist. Soc. A 175 191-216 (2012)

\bibitem{Verdery2014}
A. M. Verdery, T. Mouw, S. Bauldry, P. J. Mucha Network structure and biased variance estimation in respondent driven sampling pre-print arXiv:1309.5109 (2014)

\bibitem{Robinson2006}
W. Robinson, J. Risser, S. McGoy, A. Becker, H. Rehman, M. Jefferson, V. Griffin, M. Wolverton, S. Tortu Recruiting injection drug users: A three-site comparison of results and experiences with respondent-driven and targeted sampling procedures J. Urban Health 83 29-38 (2006)

\bibitem{McKnight2006} 
C. McKnight, D. Des Jarlais, H. Bramson, L. Tower, A. S. Abdul-Quader, C. Nemeth, D. Heckathorn Respondent-driven sampling in a study of drug users in New York City: Notes from the field J. Urban Health 83 54-59 (2006)

\bibitem{Abramovitz2009}
D. Abramovitz, E. M. Volz, S. A. Strathdee, T. L. Patterson, A. Vera, S. D. Frost, E. Proyecto Using-respondent-driven sampling in a hidden population at risk of HIV infection: Who do HIV-positive recruiters recruit Sex. Transm. Diseas. 26 (12) 750-756 (2009)

\bibitem{Iguchi2009}
M. Y. Iguchi, A. J. Ober, S. H. Berry, T. Fain, D. D. Heckathorn, P. M. Gorbach, R. Heimer, A. Kozlov, L. J. Ouellet, S. Shoptaw, W. S. Zule Simultaneous recruitment of drug users and men who have sex with men in the United States and Russia using respondent-driven sampling: Sampling methods and implications J. Urban Health 86 (1) 5-13 (2009)

\bibitem{Newman2010}
M. Newman Networks: {A}n Introduction Oxford University Press 720p (2010)

\bibitem{Costa2011}
L. F. Costa, O. N. Oliveira Jr., G. Travieso, F. A. Rodrigues, P. R. Villas Boas, L. Antiqueira, M. P. Viana, L. E. C. Rocha Analyzing and modeling real-world phenomena with complex networks: {A} survey of applications Adv. Phys. 60(3) (2011)

\bibitem{Martin2003}
J. L. Martin, J. Wiley, D. Osmond Social Networks and Unobserved Heterogeneity in Risk for AIDS Population Research and Policy Review 22 (1) 65-90 (2003)

\bibitem{Burt2010}
R. D. Burt, H. Hagan, K. Sabin, H. Thiede Evaluating respondent-driven sampling in a major metropolitan area: Comparing injection drug users in the 2005 Seattle area national HIV behavioral surveillance system survey with participants in the RAVEN and Kiwi studies Ann Epidemiol. 20 (2) 159-67 (2010)

\bibitem{McCreesh2011}
N. McCreesh, L. G. Johnston, A. Copas, P. Sonnenberg, J. Seeley, R. J. Hayes, S. D. W. Frost, R. G. White Evaluation of the role of location and distance in recruitment in respondent-driven sampling Int. J. Health Geographics 10 (56)  (2011)

\bibitem{Johnston2013}
L. G. Johnston, Y.-H. Chen, A. Silva-Santisteban, H. F. Raymond An empirical examination of respondent driven sampling design effects among HIV risk groups from studies conducted around the world AIDS Behav. 17 (6) 2202-2210 (2013)

\bibitem{Lee2006}
S. H. Lee, P.-J. Kim, H. Jeong Statistical properties of sampled networks Phys. Rev. E 73 016102 (2006)

\bibitem{Latapy2008}
M. Latapy, C. Magnien Complex network measurements: Estimating the relevance of observed properties Infocom'08 Phoenix, USA (2008)

\bibitem{Wasserman94}
S. Wasserman, K. Faust Social network analysis: {M}ethods and applications Cambridge University Press 857p (1994)

\bibitem{Gile2015}
K. J. Gile, L. G. Johnston, M. J. Salganik Diagnostics for respondent-driven sampling J. R. Statist. Soc. A (2015)

\bibitem{Serrano2005}
M. A. Serrano, M. Bogu\~na Tuning clustering in random networks with arbitrary degree distributions Phys. Rev. E 72 036133 (2012)

\bibitem{Lancichinetti2009}
A. Lancichinetti, S. Fortunato Benchmarks for testing community detection algorithms on directed and weighted graphs with overlapping communities Phys. Rev. E 80 016118 (2009)

\bibitem{Malekinejad2008}
M. Malekinejad, L. G. Johnston, C. Kendall, L. Kerr, M. R. Rifkin, G. W. Rutherford Using respondent-driven sampling methodology for hiv biological and behavioral surveillance in international settings: A systematic review  Aids Behav. 12 (4) S105-S130 (2008)

\bibitem{Lohr2009}
S. L. Lohr Sampling: Design and analysis Cengage Learning: Boston, MA 608p (2009)

\bibitem{Guimera2003}
R. Guimera, L. Danon, A. Diaz-Guilera, F. Giralt, A. Arenas Phys. Rev. E 68 065103R (2003)

\bibitem{Eckmann04}
J.-P. Eckmann, E. Moses, D. Sergi, Entropy of dialogues creates coherent structures in e-mail traffic PNAS 101 14333-14337 (2004)

\bibitem{Enron2009}
Enron Email Network http://snap.stanford.edu/data/email-Enron.html Accessed December 2014

\bibitem{Moody2001}
J. Moody Peer influence groups: Identifying dense clusters in large networks Soc. Net. 23 261-283 (2001)

\bibitem{Holme04}
P. Holme, C. R. Edling, F. Liljeros Structure and time-evolution of an {I}nternet dating community Soc. Net. 26(2) 155-174 (2004)

\bibitem{Brin1998}
S. Brin, L. Page The anatomy of a large-scale hypertextual Web search engine Comp. Net. ISDN Sys. 30(1-7) 107-117 (1998)

\bibitem{Rosvall2008}
M. Rosvall, C. T. Bergstrom Maps of random walks on complex networks reveal community structure PNAS 105(4) 1118-1123 (2008)

\bibitem{Delvenne2010}
J.-Ch. Delvenne, S. Yaliraki, M. Barahona Stability of graph communities across time scales. PNAS 107: 12755-12760 (2010)

\bibitem{Lambiotte2012}
R. Lambiotte, M. Rosvall Ranking and clustering of nodes in networks with smart teleportation. Phys. Rev. E 85: 056107 (2012)

\bibitem{Alon2007}
N. Alon, I. Benjamini, E. Lubetzky, S. Sodin Non-backtracking random walks mix faster. Commun. Contemp. Math. 9: 585 (2007)

\bibitem{Malmros2015}
J. Malmros, F. Liljeros, T. Britton Respondent-driven sampling and an unusual epidemic. Pre-print arXiv:1411.4867 (2014)

\bibitem{Newman2002}
M. E. J. Newman The spread of epidemic disease on networks. Phys. Rev. E 66: 016128 (2002)

\bibitem{Wylie2013}
J. L. Wylie, A. M. Jolly. Understanding recruitment: outcomes associated with alternate methods for seed selection in respondent driven sampling BMC Med. Res. Method. 13:93 (2013)

\bibitem{Klafter2011}
J. Klafter, I. M. Sokolov First Steps in Random Walks: From Tools to Applications. Oxford University Press, Oxford (2011)

\end{thebibliography}
\end{document}